\documentclass[10pt,letterpaper,twocolumn]{article} 

\usepackage{times,mathptmx}
\usepackage{ol2}
\usepackage{color,textcomp}
\usepackage{amsmath,amssymb}
\usepackage[english]{babel}

\begin{document}

\twocolumn[ 

\title{Dispersive wave emission and supercontinuum generation in a silicon wire waveguide
      pumped around the 1550~nm telecommunication wavelength}

\vskip-4mm

\author{Fran\c{c}ois Leo,$^{1,2,*}$ Simon-Pierre Gorza,$^3$ Jassem Safioui,$^3$ Pascal Kockaert,$^3$
        St\'ephane~Coen,$^4$ Utsav~Dave,$^{1,2}$ Bart Kuyken,$^{1,2}$ and Gunther Roelkens$^{1,2}$}

\address{
  $^1$ Photonics Research Group, Department of Information Technology, Ghent University-IMEC, Ghent B-9000, Belgium\\
  $^2$ Center for nano- and biophotonics (NB-photonics), Ghent University, Belgium\\
  $^3$ OPERA-Photonique, Universit\'e Libre de Bruxelles (ULB), 50 Av. F. D. Roosevelt, CP 194/5, B-1050 Bruxelles, Belgium\\
  $^4$ Physics Department, The University of Auckland, Private Bag 92019, Auckland 1142, New Zealand\\
  $^*$Corresponding author: francois.leo@intec.ugent.be}


\begin{abstract}%
We experimentally and numerically study dispersive wave emission, soliton fission and supercontinuum generation in a silicon wire at telecommunication wavelengths. Through dispersion engineering, we experimentally confirm a previously reported numerical study \cite{Yin_Soliton_2007} and show that the emission of resonant radiation from the solitons can lead to the generation of a supercontinuum spanning over 500\,nm. An excellent agreement with numerical simulations is observed.
\end{abstract}


 ] 



\noindent Supercontinuum generation has been the subject of many studies, particularly since the advent of photonic
crystal fibers (PCFs)~\cite{Coen_Supercontinuum_2006}. The low losses and high confinement, leading to high
nonlinearities, as well as the possibility to tailor the zero-dispersion wavelength (ZDW) has led to the generation
of supercontinua spanning over more than an octave~\cite{Ranka_Visible_2000}. Such wide spectra benefit many
applications such as high-precision frequency metrology~\cite{Jones_Carrier_2000}, optical coherence
tomography~\cite{Hartl_Ultrahigh_2001}, or telecommunication~\cite{Smirnov_Optical_2006}.

As on-chip generation of ultrashort pulses is becoming a reality~\cite{Saha_modelocking_2013,Herr_temporal_2013}, the full integration
of supercontinuum-based applications can be envisioned. On-chip supercontinuum generation was performed in
chalcogenide ~\cite{Lamont_supercontinuum_2008}, $\mathrm{Si_3N_4}$~\cite{Halir_Ultrabroadband_2012}, amorphous silicon~\cite{Safioui_Supercontinuum_2014} and silicon waveguides~\cite{Hsieh_Supercontinuum_2006,Ding_Time_2010}. On the silicon platform, previous experiments have reported a relatively limited spectral broadening in the 1550~nm telecom band. Other studies have focused instead on mid-IR pumping, and most of the broadening, up to $3.5\ \mu$m wavelength, was in that wavelength range~\cite{Kuyken_Mid_2011,Lamont_Mid_2013}. At
the 1550~nm telecom wavelength in a silicon wire, losses are not dominated by TPA but by the subsequent free carrier
absorption (FCA) and these losses typically prevent the observation of nonlinear effects~\cite{Rong_Monolithic_2006}.
Note that picosecond soliton compression has nevertheless been very recently demonstrated in silicon photonic crystals~\cite{Blanco-Redondo_2014}.
One way to circumvent the free carrier induced losses is to use very short femtosecond pulses such that the carrier density remains
negligible~\cite{Yin_Soliton_2007}. In that regime, the dynamics of supercontinuum generation is known to occur
through fission of higher-order solitons and the subsequent emission of resonant dispersive waves (DWs)
\cite{Coen_Supercontinuum_2006, Akhmediev_Cherenkov_1995, Erkintalo_Cascaded_2012, Cristiani_Dispersive_2004}. This
process is well described by the well known generalized nonlinear Schr\"odinger equation (GNLSE) and excellent
agreement between this model and experiments has been reported in PCFs~\cite{Corwin_Fundamental_2003}. Using a
similar model, Yin et al predicted that the same mechanisms should give rise to a 400\,nm wide supercontinuum in a
silicon wire at telecommunication wavelengths~\cite{Yin_Soliton_2007} but an experimental demonstration is still missing. Previous experiments reporting dispersive wave emission in silicon at telecom wavelength displayed very limited spectral broadening and the impact of the waveguide geometry was not investigated~\cite{Ding_Time_2010}.
Here we report what is, to the best of our knowledge, the first experimental observation of a broad supercontinuum by engineering the emission of dispersive waves in a silicon wire waveguide pumped in the C-band. We find an excellent agreement with simulations, confirming early predictions~\cite{Yin_Soliton_2007} that the nonlinear Schrodinger equation with a complex nonlinear parameter correctly describes broad supercontinuum generation at telecommunication wavelengths. Moreover, the good agreement with simulations gives more insight into high-order soliton dynamics in silicon in that wavelength range.

\begin{figure}[b]
  \centering
  \includegraphics[width = 8cm]{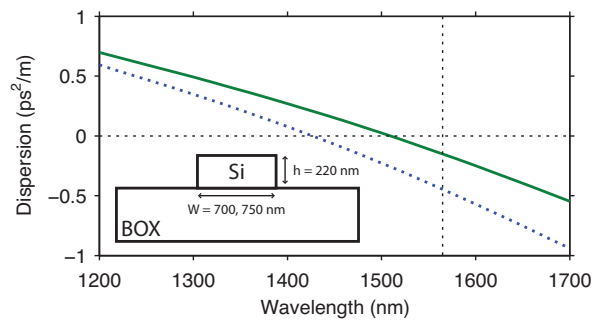}
  \caption{Simulated wavelength dependence of the second-order dispersion coefficient $\beta_2$ of 220~nm-thick SOI
    waveguides with 700\,nm (dotted) and 750\,nm (solid) width. The vertical dotted line indicates the pump wavelength.}
  \label{figdisp}
\end{figure}
We consider 7~mm-long silicon-on-insulator (SOI) waveguides with a standard 220\,nm silicon
thickness. To achieve anomalous group velocity dispersion at the pump wavelength, which is required for efficient femtosecond
supercontinuum generation~\cite{Coen_Supercontinuum_2006}, such waveguides must be no wider than 800~nm. Here we
report results for waveguides with two different widths, respectively, 700 and 750\,nm. The dispersive properties of
these waveguides have been calculated with a full vectorial mode solver, and the wavelength dependence of their
second-order dispersion coefficient $\beta_2$ is shown in Fig.~\ref{figdisp}. As can be seen, the dispersion is
small and anomalous at the 1565~nm pump wavelength used in our experiments. The pump pulses of 150~fs full-width at
half-maximum (FWHM) duration at an 82~MHz repetition rate are generated with an OPO (Spectra physics OPAL) pumped by
a Ti-Saphire laser (Spectra physics Tsunami) running at 722\,nm wavelength. We use the horizontally-polarized idler
output, exciting only the quasi-TE mode of the waveguide. The light is coupled into the waveguide with a x60
microscope objective ($\mathrm{NA}=0.65$) and coupled out with a lensed fiber ($\mathrm{NA}=0.4$). The corresponding coupling efficiencies are 17\,dB and 7\,dB.
The propagation losses are estimated to be 2\,dB/cm by cutback measurements on similar waveguides.

\begin{figure}[t]
  \centering
  \includegraphics[width = 8cm]{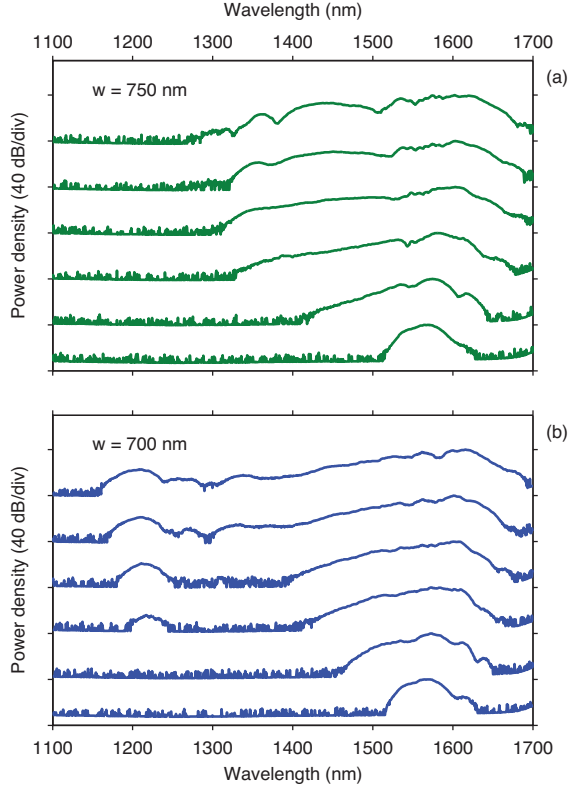}
  \caption{Experimental spectra measured at the output of (a) the 750 nm-wide and (b) the 700~nm-wide waveguide for
    on-chip peak powers of $0.15$\,W, $1.5$\,W, $3.5$\,W, 7\,W, 14\,W and 32\,W. The spectra have each been shifted
    by 40\,dB for clarity.}
  \label{figexp}
\end{figure}
The optical spectra measured at the output of the two waveguides for increasing pump peak power (up to 32\,W) are
shown in Figs.~\ref{figexp}(a) and (b). We can readily observe a clear difference between the two sets of
measurements, which reveals the strong influence of the waveguide width, hence the dispersion, on spectral
broadening in silicon wires. Resonant DWs are observed with pump peak powers as low as $3.5$\,W. They appear in the
normal dispersion regime around 1350\,nm and 1200\,nm, respectively for the 750~nm and 700~nm-wide guide.  In the
latter, the DWs allow for the generation of a supercontinuum with a -30\,dB bandwidth spanning from 1160\,nm to 1700\,nm, almost twice what was previously reported in silicon at telecom wavelength~\cite{Hsieh_Supercontinuum_2006}. We can also notice a
clear saturation of the spectral broadening when increasing the pump peak power beyond 14\,W, which can be explained
by increased two-photon absorption based nonlinear losses.

In order to gain further insights in the observed spectral broadening, we have performed numerical simulations based
on the GNLSE describing the propagation of the temporal envelope $E(z,t)$ of the electric field of short pulses
along the length~$z$ of a nonlinear medium. The equation reads~\cite{Yin_Soliton_2007, Lin_Nonlinear_2007},
\begin{multline}\label{EqNLS}
\frac{\partial E(z,t)}{\partial z} = i \sum_{k\ge 2} i^k \frac{\beta_k}{k!}\frac{\partial^k E}{\partial t^k}
 -\frac{\alpha_\mathrm{l}}{2}E -  \frac{\alpha_\mathrm{c}}{2}\left(1 + i\mu\right)E\\
+ i\gamma\left(1+\frac{i}{\omega_0}\frac{\partial }{\partial t}\right)E\int^t_{-\infty} R(t-t')|E(z,t')|^2 dt'\,.
\end{multline}
Here the $\beta_k$ are the Taylor series expansion coefficients that fit the chromatic dispersion curves of
Fig.~\ref{figdisp} in terms of angular frequency around the pump central angular frequency~$\omega_0$. 
$\alpha_\mathrm{l}$ and $\alpha_c$ account, respectively, for linear and free carrier absorption losses. We have
$\alpha_\mathrm{c} = \sigma N_\mathrm{c}$ where $N_\mathrm{c}$  is the free carrier density and $\sigma = 1.45\times
10^{-21}\,\mathrm{m}^2$ for silicon~\cite{Lin_Nonlinear_2007}. The parameter $\mu$ accounts for the free carrier
dispersion and is taken as $\mu = 2k_\mathrm{c}\omega_0/(\sigma c)$ with $c$ the speed of light in vacuum and $k_c = (8.8\times 10^{-28}N_c + 1.35\times 10^{-22}N_c^{0.8})/N_c$ ~\cite{Soref_Electrooptical_1987,Lin_Nonlinear_2007}.
Note that $k_c$ is dependent on the carrier density.
$\gamma$ is estimated from experiments on similar waveguides~\cite{Matres_High_2013} and scaled through the
effective mode area $A_\mathrm{eff}=0.2\,\mu\mathrm{m^2}$. We find $\gamma = (234 + 44 i)\,\mathrm{W^{-1}m^{-1}}$.
$R(t)$ is the nonlinearity response function defined as in fibers, $R(t) = (1-f_\mathrm{R})\delta(t) +
f_\mathrm{R}h_\mathrm{R}(t)$ where $f_\mathrm{R}$ is the fractional Raman contribution and $h_\mathrm{R}(t)$ is the
Raman response function. Both can be deduced from the known spectral Lorentzian shape of the Raman response of
silicon~\cite{Lin_Nonlinear_2007}. We use the relation $f_\mathrm{R} = g_\mathrm{R}(\omega_0)
\Gamma_\mathrm{R}/[\Omega_\mathrm{R}A_\mathrm{eff}\mathrm{Re}(\gamma)]$ with $g_\mathrm{R}(\omega_0) =3.7\times
10^{-10}\,\mathrm{m/W}$~\cite{Claps_Observation_2003}, $\Omega_\mathrm{R}/(2\pi)=15.6$\,THz and
$\Gamma_\mathrm{R}/\pi=105$\,GHz~\cite{Lin_Nonlinear_2007}  which yields $f_\mathrm{R} =0.026$. Finally, the carrier
density can be calculated by solving
\begin{equation}\label{Eqcarriers}
  \frac{\partial N_\mathrm{c}(z,t)}{\partial t} =
    \frac{2\pi \mathrm{Im}(\gamma) }{h\omega_0 A_\mathrm{eff}}|E(z,t)|^4-\frac{N_\mathrm{c}(z,t)}{\tau_\mathrm{c}}\,,
\end{equation}
where $h$ is Planck's contant and $\tau_\mathrm{c}$ is the carrier lifetime, estimated to be
1\,ns~\cite{Boyraz_All_2004}.

\begin{figure}[t]
  \centering
  \includegraphics[width = 8cm]{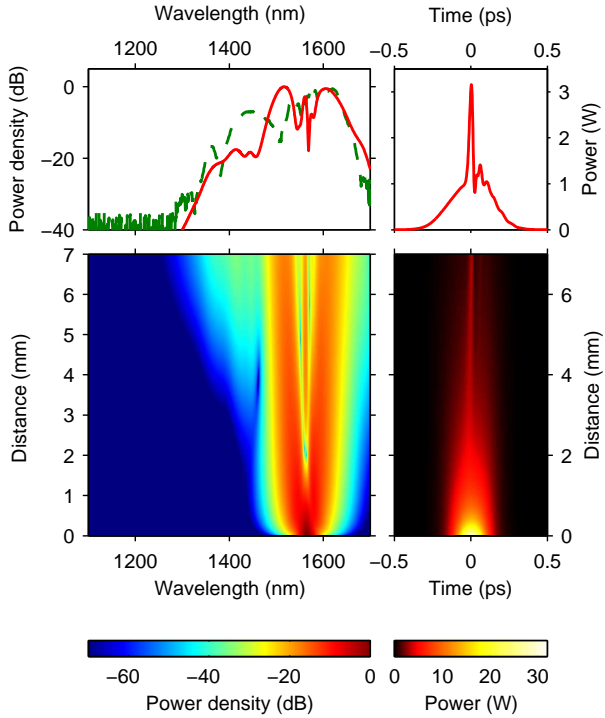}
  \caption{Pseudocolor plots of the simulated spectral (left) and temporal (right) evolution along the
    750~nm-wide silicon waveguide for a 150~fs (FWHM) 32~W sech input pulse. Top plots (red) highlight the waveguide
    output at $z=7$~mm. The dashed curve is the measured output spectrum for comparison.}
  \label{figsim750}
  \end{figure}
  \begin{figure}[t]
  \centering
  \includegraphics[width = 8cm]{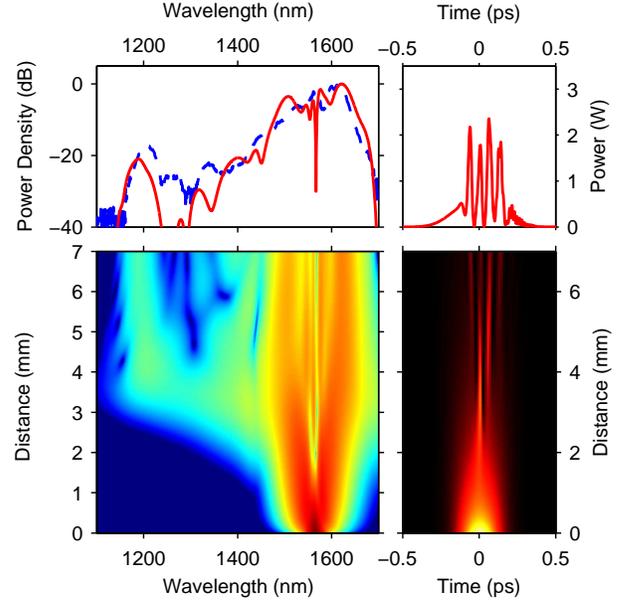}
  \caption{Same as Fig.~\ref{figsim750} but for the 700~nm-wide waveguide.}
  \label{figsim700}
\end{figure}
Equations~(\ref{EqNLS}) and (\ref{Eqcarriers}) have been solved with a split-step Fourier
algorithm~\cite{Cristiani_Dispersive_2004,Dudley_Supercontinuum_2010}. The taylor expansion  coefficients are computed up to the 10th order.
Results for a hyperbolic secant input pulse
with 150\,fs duration (FWHM) and a peak power of 32\,W corresponding to the experimental parameters are shown in
Figs.~\ref{figsim750} and \ref{figsim700} for our two different waveguides. The top-left panel of each figure
reveals a good agreement between measured and simulated output spectra. In particular the position of the DWs and
the overall spectral width are well predicted by simulations. The DWs relate to \v{C}erenkov radiation emitted by
solitons perturbed by higher-order dispersion and their spectral position can be analytically predicted with the
phase matching relation~\cite{Akhmediev_Cherenkov_1995}
\begin{equation}
  \beta(\omega_{\mathrm{DW}}) - \frac{\omega_{\mathrm{DW}}}{{v_\mathrm{g,s}}}=
    \beta(\omega_\mathrm{s})-\frac{\omega_\mathrm{s}}{v_\mathrm{g,s}} + (1-f_\mathrm{R})\gamma P_\mathrm{s},
\end{equation}			
where $\beta(\omega)$ designates the frequency-dependent wavenumber of the waveguide, $\omega_\mathrm{s}$ and
$\omega_{\mathrm{DW}}$ are the frequencies of the soliton and the emitted DW respectively, while $P_\mathrm{s}$ and
$v_\mathrm{g,s}$ are the soliton peak power and group velocity. By using a soliton peak power $P_\mathrm{s} = 10$\,W
extracted from the simulations, we find that for the 750~nm-wide (700~nm-wide) waveguide the DWs should be emitted
at 1300\,nm (1150\,nm), which is in good agreement with our experimental and numerical results. This confirms the
origin of these spectral peaks.

Additional information can be gained by examining the simulated evolution of the temporal intensity profiles of the
pulses along the waveguide, which are plotted in the right panels of Figs.~\ref{figsim750} and~\ref{figsim700}.
These figures reveal the temporal compression and subsequent splitting (or fission) of the input pulse that is
typical of supercontinuum generation in PCFs~\cite{Coen_Supercontinuum_2006}. 
The theoretical fission length \cite{Coen_Supercontinuum_2006} is 4.5\,mm for the 750\,nm-wide waveguide and 2.6\,mm for the 700\,nm waveguide which agrees reasonably well with our results despite strong nonlinear losses.
We can note however some differences with the soliton fission dynamic
reported in PCFs. Fission of a soliton of order~$N$ in PCF leads to $N$ fundamental solitons with different peak
powers and temporal widths (and these characteristics are well predicted
analytically~\cite{Coen_Supercontinuum_2006}). In contrast, while our input soliton number is $N=19$ for the
700~nm-wide waveguide, we only observe four subpulses at the output of that waveguide, and they all have roughly the
same temporal duration and peak power (note that $N=33$ for the 750\,nm-wide waveguide). We believe that the
TPA-induced peak-power limitation during the very strong temporal compression is responsible for these differences.
Previous theoretical results broadly agree with this hypothesis~\cite{Sildeberg_Solitons_1990,Afanasjev_Splitting_1995}. In particular, those works have highlighted
that TPA can induce soliton fission even in the absence of Raman scattering or higher-order dispersion, and that the
split pulses have very close characteristics. Only the cases of $N=2$ and $N=3$ have been considered in details
however, and more theoretical work is needed to fully understand the soliton fission dynamics of the GNLSE in
presence of TPA in the case of high soliton orders.
We also emphasize that our simulations predict pulse compression down to 20\,fs (see Fig.~\ref{figsim700}), highlighting the potential use of silicon for the many applications requiring ultrashort pulses.

In conclusion we have experimentally and numerically studied high-order soliton fission, dispersive wave generation,
and supercontinuum generation in a silicon photonic wire. We have reported a supercontinuum spanning from 1200\,nm
to 1700\,nm. It is obtained from 150~fs input pulses at 1565~nm wavelength, i.e., in the C-band of
telecommunications, and to the best of our knowledge it constitutes the widest reported supercontinuum in silicon in htis wavemength range.
Our work also highlights that the high-order soliton dynamics, and in particular the soliton
fission process, of the nonlinear Schr\"odinger equation is still mostly preserved in these conditions, despite the
strong influence of TPA. Let us note that we have checked that both Raman scattering and free carrier effects play
very little role in our simulations and can in practice be neglected. This results from operating in the femtosecond
regime and confirms the findings of Refs.~\cite{Yin_Soliton_2007}. We therefore believe that the good agreement
between our numerical and experimental results clearly shows that the simple NLSE with a complex nonlinear parameter
is sufficient to describe short pulse propagation in a silicon wire at telecom wavelength.

This work is supported by the Belgian Science Policy Office (BELSPO) Interuniversity Attraction Pole (IAP) programme
under grant no.~IAP-6/10 and by the FP7-ERC-MIRACLE project. The participation of S.~Coen to this project was made
possible thanks to a Research~\& Study Leave granted by The University of Auckland and to a visiting fellowship from
the FNRS (Belgium). Bart Kuyken acknowledges the special research fund of Ghent University (BOF) for a fellowship.

\end{document}